\documentclass[10pt]{article}

\usepackage{amsmath}

\usepackage{array}

\usepackage{appendix}

\usepackage{tocloft}                   

\usepackage{graphicx}

\usepackage{amsfonts}

\usepackage{amssymb}

\usepackage{mathrsfs}

\usepackage{yfonts}

\usepackage{euscript}

\usepackage{centernot}                 

\usepackage{ifsym}                     

\usepackage{upgreek}

\usepackage{mathtools}

\usepackage{color}

\usepackage{slantsc}
\usepackage{calligra}

\usepackage{bbold}          

\usepackage[T1]{fontenc}

\usepackage{epsf}

\usepackage{latexsym}

\usepackage{tipa}

\usepackage{makeidx}

\makeindex

\def\b{\begin{equation}}

\def\e{\begin{equation}}

\def\be{\begin{equation}}              

\def\ee{\end{equation}}

\def\beq{\begin{equation}}

\def\eeq{\end{equation}}

\def\bea{\begin{eqnarray}}

\def\eea{\end{eqnarray}}

\def\m{\mbox{ }}

\def\mma {\m , \m \m }

\def\!{\hspace{-1.6667em}}

\def\c{\cite}

\def\n{\noindent}

\def\u{\underline}

\def\w{\widetilde}

\def\s{\stackrel}

\def\biN{\mbox{\boldmath$N$}}

\def\mB{\mbox{B}}  

\def\mC{\mbox{C}}                        

\def\mD{\mbox{D}}                        

\def\mJ{\mbox{J}}  

\def\mK{\mbox{K}}

\def\mL{\mbox{L}}

\def\mR{\mbox{R}}                        

\def\mT{\mbox{T}}

\def\urho{{\u{\rho}}}

\def\bA{\mbox{\bf A}}

\def\bB{\mbox{\bf B}}

\def\bC{\mbox{\bf C}}                    

\def\bD{\mbox{\bf D}}

\def\bH{\mbox{\bf H}}

\def\bI{\mbox{\bf I}}

\def\bJ{\mbox{\bf J}}

\def\bK{\mbox{\bf K}}

\def\bL{\mbox{\bf L}}

\def\bQ{\mbox{\bf Q}}

\def\bR{\mbox{\bf R}}

\def\bSigma{\mbox{\boldmath$\Sigma$}}                 
 
\def\bLambda{\mbox{\boldmath$\Lambda$}}                 

\def\ia{\mbox{\normalsize $a$}}

\def\sT{\mbox{\scriptsize T}}

\def\sumi2{\sum\mbox{}_{\mbox{}_{\mbox{\scriptsize $i$=1}}}^2}

\def\sumi3{\sum\mbox{}_{\mbox{}_{\mbox{\scriptsize $i$=1}}}^3}

\def\sumABcycles3{\sum\mbox{}_{\mbox{}_{\mbox{\scriptsize cycles $A,B$=1}}}^{3}}

\def\sumCDcycles3{\sum\mbox{}_{\mbox{}_{\mbox{\scriptsize cycles $C,D$=1}}}^{3}}

\def\sumj3{\sum\mbox{}_{\mbox{}_{\mbox{\scriptsize $j$=1}}}^3}

\def\sumk3{\sum\mbox{}_{\mbox{}_{\mbox{\scriptsize $k$=1}}}^3}

\def\prodiA1{\prod\mbox{}_{\mbox{}_{\mbox{\scriptsize $i$=1}}}^{A - 1}}

\def\bigtimes{\mbox{\Large $\times$}}

\def\es{\m = \m}

\def\:={\m := \m}

\def\=:{\m =: \m}

\def\FrS{\mbox{\Large $\mathfrak{s}$}}                         
\def\FrW{\mbox{$\mathfrak{W}$}}                                

\def\Hilb{\mbox{{\boldmath$\mathfrak{H}$}ilb}}                 
\def\FrQ{\mbox{\Large $\mathfrak{q}$}}                               
\def\Phase{\mbox{{\boldmath$\mathfrak{P}$}hase}}                     

\def\bFrR{\mbox{\boldmath$\mathfrak{R}$}}                            
\def\Rig-Phase{\bFrR\mbox{ig-}\Phase}                                
                                                                													   
\def\lFrr{\mbox{\Large $\mathfrak{r}$}}                              

\def\FrP{\mbox{\Large $\mathfrak{p}$}}                                 
\def\FrR{\mbox{\boldmath$\mathfrak{R}$}}                             

\def\bFrM{\mbox{\boldmath${\mathfrak{M}}$}}                             

\def\bFrR{\mbox{\boldmath$\mathfrak{R}$}}                            

\def\bFrR{\mbox{\boldmath$\mathfrak{R}$}}                            

\def\1mat{\u{\u{1}}}                                                 

\def\Positive-Modespace{\mbox{{\boldmath$\mathfrak{M}$}odespace$^+$}}

\def\POSITIVE-MODESPACE{\mbox{{\boldmath$\mathfrak{M}$}ODESPACE$^+$}}

\def\FrO{\mbox{$\mathfrak{O}$}}                                      

\def\Kin-Hilb{\mbox{{\boldmath$\mathfrak{K}$}in-\Hilb}}                     

\def\Mid-Hilb{\mbox{{\boldmath$\mathfrak{M}$}id-\Hilb}}                     

\def\Dyn-Hilb{\mbox{{\boldmath$\mathfrak{D}$}yn-\Hilb}}                     

\def\5Star{\mbox{\Large$\star$}}              

\begin{document}

\begin{titlepage}

\begin{center}

\Large{\bf Isotropy Groups and Kinematic Orbits}

\vspace{0.1in}

{\bf for 1 and 2-$d$ $N$-Body Problems} \normalsize

\vspace{0.2in}

{\normalsize \bf Edward Anderson$^*$}

\vspace{.2in}

\end{center}

\begin{abstract}
 
\n Mitchell and Littlejohn showed that isotropy groups and orbits for $N$-body problems attain a sense of genericity for $N = 5$. 
The author recently showed that the arbitrary-$d$ generalization of this 3-$d$ result is that genericity in this sense occurs for $N = d + 2$.  
The author also showed that a second sense of genericity -- now order-theoretic rather than a matter of counting -- occurs for $N = 2 d + 1$, 
excepting $d = 3$, for which it is not 7 but 8.  
Applications of this work include 1) that some of the increase in complexity in passing from 3 to 4 and 5 body problems in 3-$d$ is already present in the more-well known setting 
of passing from intervals to triangles and then to quadrilaterals in 2-$d$. 
2) That not $(d, N) = (3, 6)$ but (4, 6) is a natural theoretical successor of (3, 5).
3) Such consideration isotropy groups and orbits is moreover a model for a larger case of interest, namely that of GR's reduced configuration spaces.  
The current Article presents the lower-$d$ cases explicitly:  0, 1 and 2-$d$, including also the topological and geometrical form of the corresponding isotropy groups and orbits. 

\end{abstract}

\n {\bf PACS}: 04.20.Cv , 02.40.Yy , 02.70.Ns .

\m 

\n {\bf Physics keywords}: $N$-Body Problem, Configuration Spaces, Background Independence, Topological and Geometrical Methods in Theoretical Physics.

\m

\n {\bf Mathematics keywords}: Shape Theory, Applied Geometry, Topology and Linear Algebra. 

\vspace{0.1in}
  
\n $^*$ Dr.E.Anderson.Maths.Physics *at* protonmail.com

\section{Introduction}\label{Auts}

\n The $N$-Body Problem rapidly develops technical complexity with increasing $N$ \cite{McGehee, MM75, Palmore, ABC96, LR97, LMRAC98, Albouy-1, Xia, ML00, Albouy, Minimal-N}.  
Even $N = 3$ is considered to be hard, $N = 4$ harder, and $N = 5$ the limit of current-era detailed study for specific $N$. 
This is moreover with reference to 3-$d$; 1- and 2-$d$ exhibit a number of more systematic features \cite{Moulton, Smale70, McGehee, MM75, Kendall84, Kendall, Minimal-N}
Finally, we recently noted \cite{Minimal-N} that many qualitative features depend on the combination $(d, \, N)$ rather than just on $d$, 
placing further interest in lower-$d$ specific examples (the current Article) and higher-$d$ ones (for subsequent study: see the Conclusion).  

\m 

\n Our study is at the level of configuration space, more specifically of shape space and relational space (alias shape-and-scale space).  
These are outlined in Sec 2, and constitute an elementary part of Shape Theory \cite{Kendall84, Kendall89, Small, Sparr, Kendall, MP05, GT09, FileR, Bhatta, QuadI, DM16, PE16, KKH16, 
ABook, I, II, III, III-Extended}.
These structures arise moreover as part of Background Independence \cite{A64, Dirac, A67, DeWitt67, Battelle, BB82, HT92, Giu06, APoT3, ASoS, ABook, PE-1, AObs4};  
the well-known Problem of Time \cite{Battelle, DeWitt67, K92, I93, APoT, APoT2, APoT3, ABook} can moreover be viewed as difficulties arising in attempting to 
implement Background Independence. 

\m

\n The current Article considers more specifically isotropy groups and kinematic orbits in the shapes-and-scales case of quotienting out by the Euclidean group.
We approach this with various uses of Linear Algebra and one of Differential Geometry in Sec 3, and explicit examples in Sec 2.
By further material in Sec 2, this readily reduces to just quotienting out rotations by passing to centre of mass frame and applying the Jacobi map. 
We consider, firstly, Mitchell and Littlejohn's criterion \cite{ML00} of minimal point-or-particle number $N$ required to distinctly realizing a full count of isotropy subgroups.
Secondly, which minimal $N$ is required for these isotropy subgroups to form the generic bounded lattice of subgroups \cite{Minimal-N}.
These are `C' and `O' genericity criteria respectively, standing for `counting' and `order'.  
We concentrate on the 0-, 1- and 2-$d$ cases, \cite{ABC96, LMRAC98, ML00, Minimal-N} having already detailed 3-$d$, 
working as far up as the first counting-generic and order-generic $N$.
Key underlying results are, firstly, that C-genericity occurs for 
\be 
N = d + 2
\ee 

\end{titlepage}

\n rather than just specifically for $(d, \, N) =(3, \, 5)$.  
Secondly, that a bound on O-genericity occurs for 
\be 
N = 2 \, d + 1
\ee 
except for $d = 3$ itself, for which a Lie group accidental relation pushes it up to 8.  
This means that the well-studied quadrilaterals (2, 4) are a model arena for (3, 5), and that the pentagons (2, 5) are in some ways a model arena for the O-generic (3, 8).  
The well-known progression from intervals (2, 2) to triangles (2, 3) to guadrilaterals 
is moreover a model for some features of the key progression in complexity in passing from 3- to 4- to 5-Body Problems in 3-$d$.  

\m  
 
\n Such consideration isotropy groups and orbits is moreover of further interest as a model for a larger case of interest, 
namely that of GR's reduced configuration spaces \cite{DeWitt67, DeWitt70, Fischer70, FM96, Giu09}, 
as well as of some specifically global \cite{ABook} aspects of Background Independence and the Problem of Time.   

\m 

\n In the current Article, we provide 0, 1 and 2-$d$ counterparts of \cite{ML00}'s analysis in support of \cite{Minimal-N}'s 
point that their analysis is counting-generic for . 
Thus (0, 2), (1, 3) and (2, 4) are considered here, as well as the partial realizations for smaller $N$'s than these in each case. 
We conclude with an outline of pointers to  further research directions in Sec 8; see also \cite{Minimal-N, Minimal-N-2} for further details.

\section{Some configuration spaces for \biN-Body Problem}

\n The {\it carrier space} $\bFrM^d$, 
                   alias {\it absolute space} in the physical context is an at least provisional model for the structure of space.

\m

\n{\it Constellation space} (reviewed in \cite{FileR}) is the product of $N$ copies of carrier space, 
\be 
\FrQ(\bFrM^d, \, N) \es \bigtimes_{I = 1}^N \bFrM^N
\ee 
modelling $N$ points on $\bFrM$, or, if these points are materially realized, $N$ particles (classical, nonrelativistic). 

\m 

\n In the current Article, we consider the most common setting for the $N$-Body Problem, 
\be 
\bFrM = \mathbb{R}^n \m , 
\ee
by which 
\be 
\FrQ(d, \, N)  :=   \FrQ(\mathbb{R}^d, \, N)  
               \es  \bigtimes_{I = 1}^N \mathbb{R}^{d}  
			   \es  \mathbb{R}^{N \, d}                               \m .
\ee
We further narrow this down to $d = 0, 1, 2$, \cite{ML00} having already carried out the analysis in question for the most commonly considered $N$-body setting of all, 
$\mathbb{R}^3= 3$.  

\m 

\n{\it Relative space} (see \cite{FileR, I, Minimal-N})is the quotient of constellation space by the group of translations, 
\be 
Tr(d)  =  \mathbb{R}^d                                                \m ,  
\ee 
yielding 
\be 
\lFrr(d, \, N)  \:=  \frac{\FrQ(\mathbb{R}^d, \, N)}{Tr(d)}  
                \es  \frac{\mathbb{R}^{d \, N}}{\mathbb{R}^d}
                \es  \mathbb{R}^{d \, n}			                  \m .
\ee
This last equation holds both topologically and metrically, 
as is most easily seen \cite{Minimal-N} by passing to mass-weighted relative Jacobi coordinates \cite{Marchal}.  
These maintain form of kinetic metric but with one object less; this can be viewed as result of diagonalizing the relative separations. 
\be 
n  :=  N - 1
\ee
is the number of independent relative separations.  
 
\m 

\n{\it Preshape space} \cite{Kendall84, Kendall, I} is the quotient of constellation space by the {\it dilatational group} comprised of translations and dilations, 
\be 
Dilatat(d)  \es  Tr(d) \rtimes Dil 
            \es  \mathbb{R}^n \rtimes \mathbb{R}_+                                                            \m , 
\ee
where $\rtimes$ denotes semidirect product of groups \cite{Cohn}.  
This yields 
\be 
\FrP(d, \, N)  \:=  \frac{\FrQ(d, \, N)}{Dilatat(d)} 
               \es  \frac{\mathbb{R}^{N \, d}}{\mathbb{R}^n \rtimes \mathbb{R}_+      }
               \es  \frac{\mathbb{R}^{d \, N}}{\mathbb{R}_+}			   
			   \es  \mathbb{S}^{n \, d - 1}                                                                   \m , 
\ee
this last result being Kendall's preshape sphere, at both the topological and metric levels of structure. 

\m 

\n{\it Shape space} \cite{Kendall84, Kendall89, Kendall} is the quotient of constellation space by the {\it similarity group} comprised of translations, dilations and rotations, 
\be 
Sim(d)  \es  Tr(d) \rtimes (Dil \times SO(d)) 
        \es  \mathbb{R}^n \rtimes (\mathbb{R}_+ \times SO(d))                                                 \m , 
\ee
where $\times$ denotes direct product of groups \cite{Cohn}.  
This yields 
\be 
\FrS(d, \, N)  \:=  \frac{\FrQ(d, \, N)}{Sim(d)} 
               \es  \frac{\mathbb{R}^{N \, d}}{\mathbb{R}^n \rtimes (\mathbb{R}_+ \times SO(d) )      } 
               \es  \frac{\mathbb{R}^{n \, d}}{\mathbb{R}_+ \times \times SO(d)      } 
			   \es  \frac{  \mathbb{S}^{n \, d - 1}  }{  SO(d)  }                                             \m . 
\ee
\n In 1-$d$, there are no continuous rotations, so shape space coincides with preshape space
\be 
\FrS(1, \, N) = \FrP(1, \, N) = \mathbb{S}^{n - 1}                                                            \m .  
\ee
\m 
In 2-$d$, 
\be 
SO(2) = U(1)
\ee 
and e.g.\ \cite{A-Coolidge} the generalized Hopf map \cite{Nakahara} gives that 
\be 
\FrS(2, \, N)  \es  \frac{  \mathbb{S}^{2 \, n - 1}  }{  U(1)  } 
               \es  \mathbb{CP}^{n - 1}                                                                       \m :
\ee 
complex-projective spaces of $N$-a-gons \cite{Kendall84, Kendall, QuadI} as equipped with the standard Fubini--Study metric \cite{Nakahara}.

\m 

\n Exceptionally for $N = 3$, the Hopf map itself gives
\be 
\FrS(2, \, 3)  \es \mathbb{CP}^{1} 
               \es  \mathbb{S}^{2}      
\ee 
-- the sphere of triangles in 2-$d$ \cite{Kendall84, Kendall89, Kendall, III} --with this case's Fubini--Study metric collapsing to the standard spherical metric.  

\m 

\n{\it Relational space} alias {\it scaled shape space} is the quotient of constellation space by the {\it Euclidean group} of translations and rotations, 
\be 
Eucl(d)  \es  Tr(d) \rtimes (Dil \times SO(d)) 
         \es  \mathbb{R}^n \rtimes \times SO(d)                \m . 
\ee
This yields 
\be 
\FrR(d, \, N)  \:=  \frac{\FrQ(d, \, N)}{Eucl(d)} 
               \es  \frac{\mathbb{R}^{N \, d}}{\mathbb{R}^n \rtimes SO(d) } 
			   \es  \frac{  \mathbb{R}^{n \, d}  }{  SO(d)  }
               \es	\mC(\FrS(d, \, N))		                                                              \m ,  
\ee
where $C( \FrW  )$ denotes the topological and metric-level cone \cite{FileR} over the space $\FrW$.
\n In 1-$d$, this simplifies to 
\be 
\FrR(1, \, N)  =  \mC(\mathbb{S}^{n - 1}) 
               =  \mathbb{R}^n
               =  \lFrr(1, \, N)                              \m : 
\ee
the flat relative space once again.  

\m 

\n In 2-$d$, however \cite{QuadI}, 
\be 
{\cal R}(2, \, N)  =  \mC(\mathbb{CP}^{n - 1})  
\ee
neither coincides with relative space nor elsewise simplifies.  

\m  

\n Finally, in 0-$d$, all the points pile up, and none of translations, dilations or rotations are defined, 
so the trivial configuration space of piled-up points suffices for {\sl all} of the above spaces. 

\vspace{10in}

\section{Further orbit and isotropy group structure}\label{main}

\n The {\it kinematic group} of the $N$-Body Problem in $\mathbb{R}^d$ is the `internal' $SO(n)$ rotations 
acting on whichever $\lFrr(d, N)$ basis choice of $n$ mass-weighted Jacobi vectors $\u{\rho}^i$ in the natural manner. 
This treats the components of each $\u{\rho}^i$ together as a package.

\m 

\n A {\it kinematic rotation} alias {\it internal rotation} \cite{ML00} $\bK \, \in \, SO(n)$ is one which acts `internally' 
-- i.e.\ not at the level of space but of configuration space, more concretely relative space $\lFrr(d, \, n)$ --
by exchanging the relative-separation-cluster labels $i$ of the Jacobi vectors $\u{\rho}^i$ according to the linear combination
\be 
\u{\rho}^i \m \longrightarrow \m \sum_{j = 1}^{n}{K^i}_j \u{\rho}^j
\ee 
Note that $N = 1$ has $n = 0$ and so no $\urho^i$ for any kinematical rotations to act upon. 
$N = 2$ is then minimal for kinematical rotation matrices to be defined, though its $SO(1)$ can just be  the identity rather than a nontrivial linear combination.
$N = 3$ is thus the minimal requirement for a nontrivial kinematical rotation group in the sense of admitting nontrivial linear combinations of Jacobi vectors.

\m 

\n For 2 or 3-$d$, we also have an arbitrary {\it external} alias {\it spatial rotation} $\bL \, \in \, SO(d)$.
These obey 
\be
\mbox{\bf [}\bK \mbox{\bf ,} \, \bL \mbox{\bf ]}  =  0              \m , 
\ee 
by which \cite{ML00} $SO(n)$ has a well-defined action on the relational space $\FrR(d, \, N)$.
In 2-$d$, there is a single scalar $\mL$, whereas in 1-$d$  $SO(d)$ rotations collapse to just the identity and so can be omitted from the analysis.

\m 

\n We next form an array $\bJ$ with components 
\be 
J^{ia} = \rho^{ia} \, 
\ee  
so it is built by adjoining $n$ Jacobi vector columns of height $d$ each to form a $d \times n$ array.  
This transforms according to 
\be 
\bJ = \bL \, \bJ         \m , 
\ee 
\be 
\bJ = \bJ \, \bK^{\sT}   \m . 
\ee 
\n The {\it kinematic group orbit} through a specific relative configuration $\bJ$ is 
\be 
\FrO(\bJ) :=  \{ \, \bK \, \mR \, | \, \bK \, \in \, SO(n) \, \}                                          \m .
\ee
Let us also define the {\it isotropy group} alias {\it stabilizer} corresponding to our kinematical action on $\bJ$ by  
\be 
Isot(\bJ)  :=  \{ \, \bK \, \in \, SO(n) \, | \, \bJ \m \mbox{ is invariant under } \bK  \, \in \, SO(d) \, \} 
            =  \{ \, \bK \, \in \, SO(n) \, | \, \bJ \, \bK^{\sT} = \bQ \,  \bJ \, \}               \m . 
\label{Isot}
\ee 
The last equality here follows from $\bJ \, \bK^{\mT}$ and $\bJ$ having the same scaled shape iff they are related by a spatial rotation $\bL$.

\m 

\n To find the isotropy subgroups, we proceed via the following Linear Algebra treatment (\cite{ML00} but in arbitrary dimension).  
$\bJ$ furthermore admits a {\it principal value decomposition}  
\be 
\bJ = \bL \, \bLambda \, \bH^{\sT} 
\ee 
for $\bL \, \in \, SO(d)$, $\bH \, \in \, SO(n)$ and 
\be 
\Lambda = (\mbox{diag}(\lambda_i), \, \u{0}, \, . . . \, \u{0})
\ee 
with $n - d$ zero columns, so that this is overall a $d \times n$ array.  

\m

\n We next recast the second equality in (\ref{Isot})'s condition as 
\be 
\mbox{for which } \m \bK \, \in \, SO(n) \m \exists \m \bL \, \in \, SO(d) \m \mbox{ such that } \m \bL \, \bLambda \, (\bH^{\sT} \bK \, \bH)^{\sT} = \bLambda.
\label{Isot-2}
\ee 
This is attained by two uses of our principal decomposition followed by some basic properties of rotations, transposes and inverses.
(\ref{Isot-2}) signifies that the isotropy subgroup of the kinematic group's action on the scaled shape $\bR$ is (group-theoretically) conjugate \cite{Cohn} to 
                            the isotropy subgroup of the kinematic group's action on the scaled shape $\bLambda$ 
This is a simple (and thus useful) way of representing that $Isot(\bJ)$ is moreover really a function just of scaled shape $\bR$, $Isot(\bR)$,  
from $\bJ$ already being translation-invariant and definition (\ref{Isot})  evoking $SO(d)$-invariance.

\m 

\n We next note that \cite{ML00}, at the Differential-Geometric level, 
\be 
\FrO(\bR) \m \s{\mbox{diffeo}}{=} \m  \frac{SO(n)}{Isot(\bR)}                                         \m .   
\ee 							
Replacing the subgroup being quotiented out by a conjugate subgroup leaves the resulting orbit manifold $\FrO$ invariant (up to diffeomorphism). 
Following \cite{ML00}, we thus assume without loss of generality that
\be 
\bJ \m \mbox{ is represented by } \m \bLambda  \m .
\ee 
We are thus seeking all $K \, \in \, SO(n)$ such $\exists$ some $\bL \, \in \, SO(d)$ satisfying 
\be 
\bL \, \bLambda  \, \bK^{\sT} = \bLambda \m . 
\label{Key}
\ee 
\be 
r := \mbox{rank}(\bLambda)  \m 
\ee 
subsequently plays a significant role in this calculation, receiving moreover the geometrical interpretation of the dimensionality of the scaled shape $\bR$ in question.  
I.e.\ 
\be 
r = 0  \m \mbox{ are maximal coincidences-or-collisions of points-or-particles}  \m , 
\ee 
\be 
r = 1 \m \mbox{ are collinear shapes}                                            \m , 
\ee 
\be 
r = 2 \m \mbox{ are coplanar shapes}                                             \m . \m . \m .
\ee 
$\bLambda$ can furthermore in all cases be written in the block form  
\be
\bLambda = \mbox{\Huge(} \s{\mbox{$\bSigma$}}{0} \m \s{\mbox{0}}{0} \mbox{\Huge)}
\label{Rank-Block}
\ee 
for 
\be
\bSigma = \mbox{diag}(\lambda_1, \, ... \, , \, \lambda_r)  \m . 
\ee 
This is by changing basis to place $\bLambda$'s zero eigenvalues last, 
so they are contiguous with the zero column vectors added on to the right of $\bLambda$ in defining this array. 
One can then `redecompose' into $r \times r$, $r \times (n - r)$, $(d - r) \times r$ and $(d - r) \times (n - r)$ blocks as per (\ref{Rank-Block}).  

\m  

\n(\ref{Key}) moreover only admits solutions if $\bK$ and $\bL$ are themselves in block-diagonal form, which we denote by 
\be 
\bK = \mbox{\Huge(} \s{\mbox{$\bA$}}{0} \m \s{\mbox{0}}{\bB} \mbox{\Huge)} \m \mbox{ and } 
\ee 
\be 
\bL = \mbox{\Huge(} \s{\bC}{0} \m \s{\mbox{0}}{\bD} \mbox{\Huge)} \m 
\ee 
respectively.  
$\bA$, $\bB$, $\bC$, $\bD$ here are $r \times r$, $(n - r) \times (n - r)$, $r \times r$ and $(d - r) \times (d - r)$ matrices respectively.  
Solving for our problem's $\bK$ and $\bL$ is furthermore equivalent to finding orthogonal matrices $\bA$, $\bB$, $\bC$, $\bD$ satisfying the algebraic system of equations 
\be 
\bC \, \bSigma \, \bA^{\sT} = \bSigma              \m , 
\label{CSAS}
\ee 
\be 
\mbox{det} \, \bA \, \mbox{det} \, \bB = 1       \m , 
\ee
\be 
\mbox{det} \, \bC \, \mbox{det} \, \bD = 1       \m , 
\label{CD1}
\ee 
Finally, since $\bSigma$ is invertible and $\bA$, $\bD$ are orthogonal, (\ref{CSAS}) requires that 
\be 
\bC = \bA                                        \m . 
\ee 
(\ref{CD1}) thus becomes 
\be 
\mbox{det} \, \bA \, \mbox{det} \, \bD = 1   \m .
\ee 
Moreover, on the one hand, for maximal rank $r = d$, $\bD$ becomes a zero-dimensional block so 
\be 
\mbox{det} \, \bA = 1 \m , 
\label{det1}
\ee 
which just says that $\bA$ must be special-orthogonal. 
On the other hand, for non-maximal rank, we can always find a $\bD$ of equal-sign determinant to $\bA$, so the last equation is `ineffective' 
(in the sense of not imposing any new restrictions).

\section{Minimal-$N$ cases dimension by dimension}

\n \cite{ML00} also noted that for small enough $N$, not all isotropy subgroups are distinctly realized. 
Avoiding this singled out $N = 5$ for their 3-$d$ analysis as the minimal case that attains genericity in this sense. 
We moreover showed that \cite{Minimal-N} this is in fact not a property of 5-body problems but of the interplay between $d$ and $N$, 
occurring specifically for $N = d + 2$ in dimension $d$.
Thus some of the increase in complexity in the 3, 4, 5 body problem sequence in 3-$d$ is paralleled by that in the more familiar case of passing from 
the intervals to the triangles to the quadrilaterals in 2-$d$.  
[The 1-$d$ parallel, from 1 to 2 and 3 points in 1-$d$, is probably too inherently simple as an analogy; see also the Conclusion for an outline of further $d \geq 4$ parallels.]  
Since we also argued for a further notion of genericity in \cite{Minimal-N}, 
let us qualify the above use of genericity moreover as {\it C-genericity}, since it corresponds to having a full {\sl count} of distinctly realized isotropy groups.  

\m 

\n The first main result of the current Article is as follows.  

\m  

\n{\bf Proposition 1} $d = 0, 1, 2, 3$ possess 1, 2, 4 and 7 isotropy groups respectively as per Fig \ref{7-Isot}, 
                                            with a corresponding number of kinematical orbits as per \ref{7-Orb}.  
$N = 2, 3, 4$ and 5 are minimal to distinctly realize these in 0- to 3$d$ respectively.  

\m 

\n The 3-$d$ case of this was worked out by Mitchell and Littlejohn \cite{ML00}, 
building on earlier work with Reinsch, Aquilanti and Cavalli \cite{LMRAC98} for the $N = 4$ model.  
The current Article provides the corresponding workings for 0- to 2-$d$. 

\m

\n{\bf Proposition 2} $N = 2$, 3, 5 and 8 are minimal to distinctly realize the continuous parts of the isotropy groups in 0- to 3-$d$ respectively.

\m  

\n This is derived in \cite{Minimal-N} for 3-$d$ and in the current Article for 0- to 2-$d$.  
%
{\begin{figure}[ht]
\centering
\includegraphics[width=0.8\textwidth]{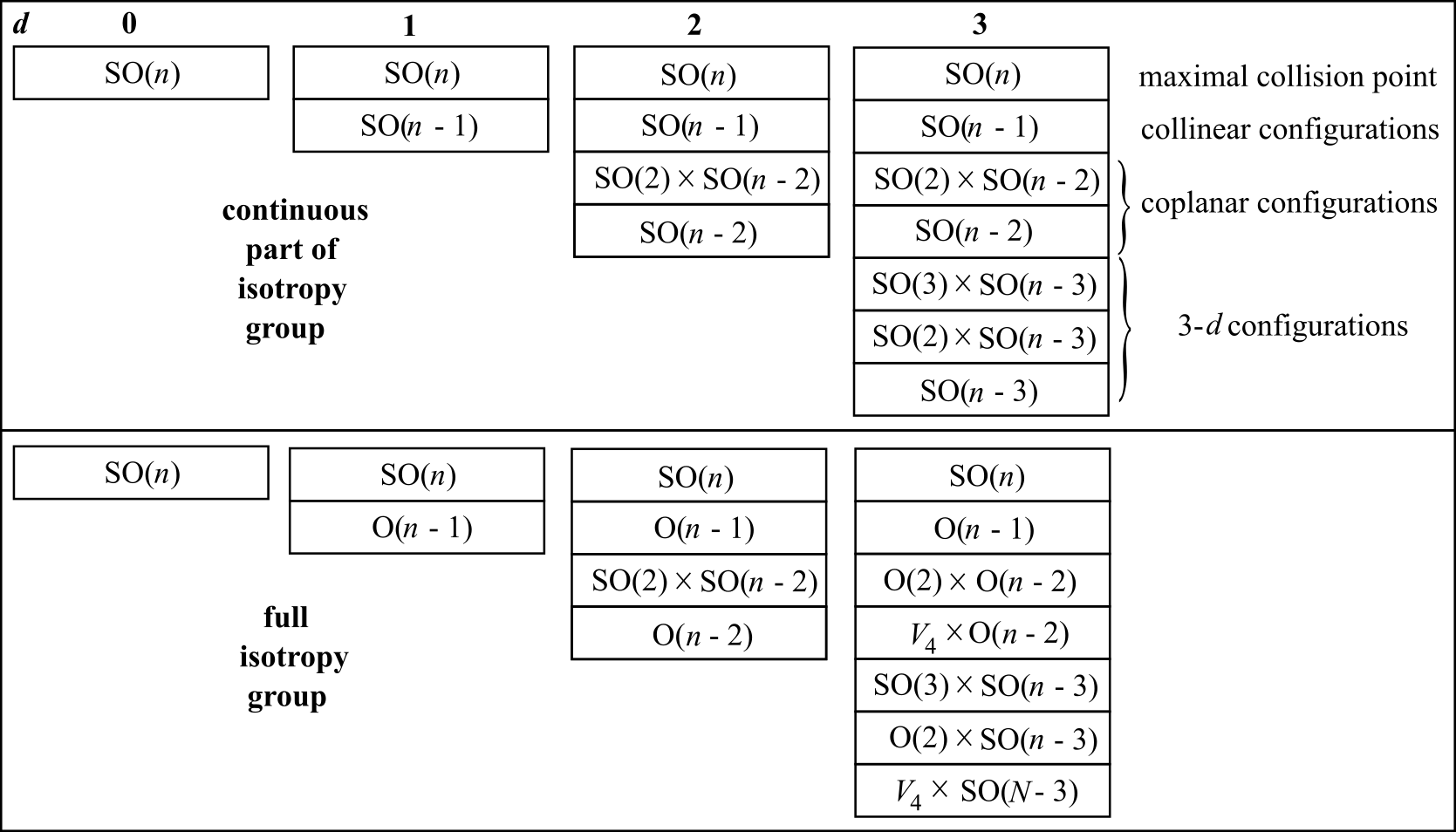}
\caption[Text der im Bilderverzeichnis auftaucht]{\footnotesize{Isotropy groups in a)1-$d$, b) 2-$d$, and c) in 3-$d$. 
The first row are the continuous parts of the group, whereas the second row are the full group: discrete parts included.
$V_4 := C_2 \times C_2$: the Klein 4-group.}} 
\label{7-Isot}\end{figure} } 
%
{\begin{figure}[ht]
\centering
\includegraphics[width=1.0\textwidth]{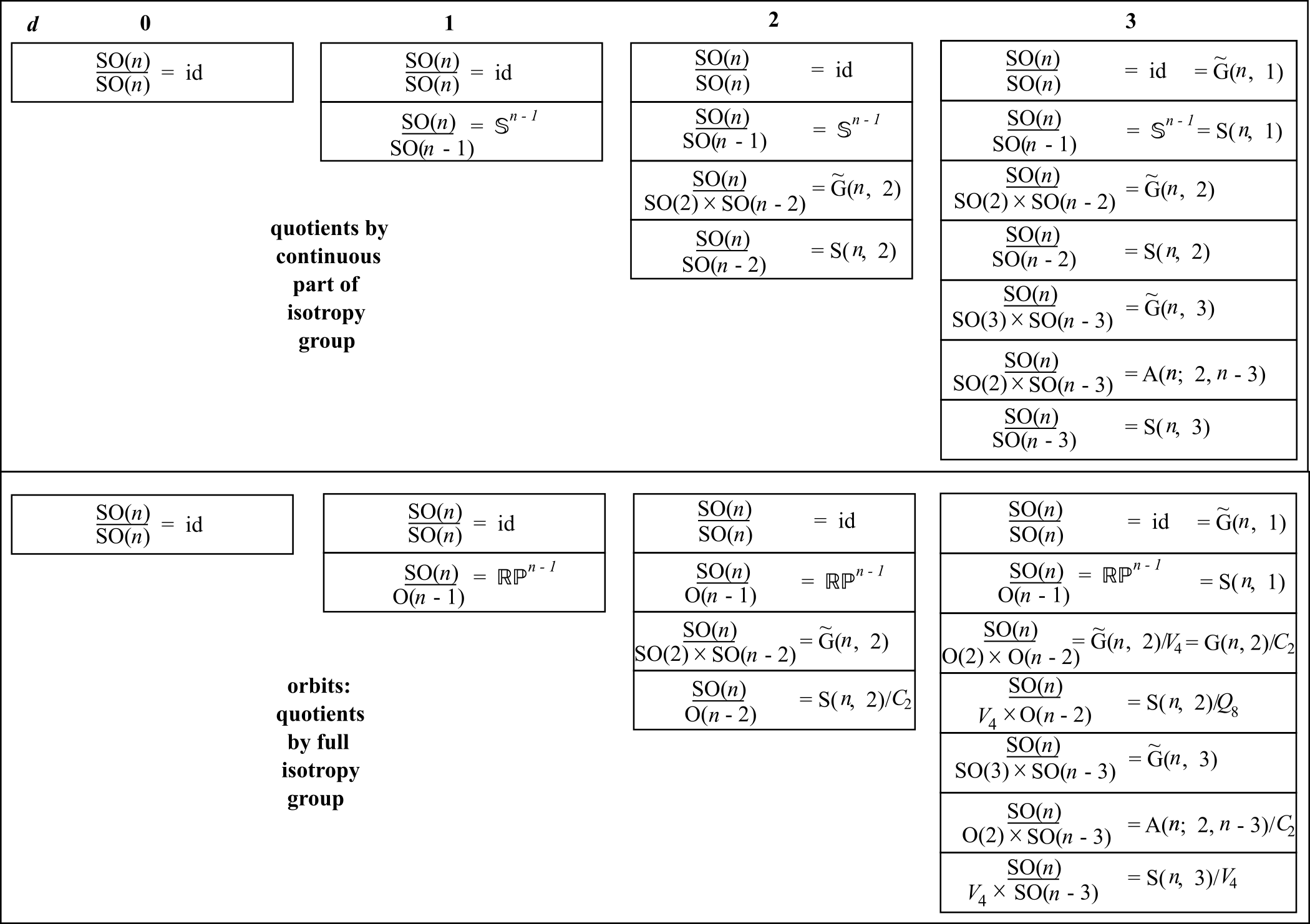}
\caption[Text der im Bilderverzeichnis auftaucht]{\footnotesize{The corresponding orbit space topologies.
$Q_8$ is the discrete order-8 group of quaternions. 
See (\ref{Stiefel}-\ref{A-Space}) for what $S$, $\w{G}$, $G$ and $A$ are.  
}} 
\label{7-Orb}\end{figure} } 

\m  

\n A generic chain of isotropy subgroups of length $d$ is present in each dimension $d$.  
To introduce our notation and explain the ensuing geometry, on the one hand, the bottom element of each such chain is a {\it real Stiefel space} 
(see \cite{NSBook, Nakahara, AMP2, Frankel} for introductory accounts or \cite{Husemoller} for a more detailed account), 
\be 
S(p, q) \:= \frac{SO(p)}{SO(q)}                                                                                     \m .  
\label{Stiefel}
\ee 
On the other hand, each bottom element is a {\it real oriented Grassmann space} \cite{NSBook, Nakahara, AMP2, Frankel, Husemoller}
\be 
\w{G}(n, d)  \es  \frac{SO(n)}{SO(d) \times SO(n - d)}                                                              \m .  
\label{Grass-tilde}
\ee
{\it Unoriented real Grassmann spaces} 
\be 
G(n, d)  \es  \frac{O(n)}{O(d) \times O(n - d)}                                                                     \m .  
\label{Grass}
\ee
moreover also feature in Fig 2's list of cases of relevance. 

\m 
 
\n Finally, the middle elements of the continuous-part chain are moreover more general than Grassmann spaces, along the lines of  
\be 
A(p; q, r) =  \frac{SO(p)}{SO(q) \times SO(r)}  \mma q + r \neq p                                                   \m .   
\label{A-Space}
\ee 
These `A-spaces' are outlined in \cite{Minimal-N}, 
but are not required for the current Article's specific examples since for $d \leq 2$ there is no room for our chains to contain nonextremal elements.

\m 

\n{\bf Proposition 3} The general result is that 
\be 
N - 1 = n = d + 1   \m \mbox{ is minimal for C-genericity}                 \m .   
\ee 
Thus \cite{Minimal-N}
\be 
\mbox{(minimal linear dependence)} \m = \m \mbox{(minimal C-genericity)}   \m . 
\ee 
\n{\u{Derivation}} This follows by generalizing Mitchell and Littlejohn's point about the smallest-dimension generic orbit, which in their $d = 3$ case has dimension 
\be 
3 \, N - 12  =  3 ( N - 4 )
\ee 
thus requiring $N \geq 5$ to realize.
For such dimension counting, the continuous part of the orbit suffices. 

\m 

\n From the second factor in this, these are all zero-dimensional along the basis diagonal. 
Inclusion of one more point-or-particle than the basis diagonal however suffices for this to attain a positive-integer value.  $\Box$

\m 

\n{\bf Remark 4} Once variable dimension is incorporated, 
Mitchell and Littlejohn's condition is not a bound on $N$ but rather a ladder of unit slope in the $(d, \, N)$ grid, with $N = 5$ being the $d = 3$ case. 

\m 

\n On the one hand, the quadrilaterals in the plane           -- $(d, N) = (2, 4)$ -- are revealed to be a meaningful model arena for 
the notoriously hard and interesting 5-Body Problem in 3-$d$:    $(d, N) = (3, 5)$, 
with the                          step-up in complexity from the tetrahaedrons' (3, 4) to (3, 5) sharing some conceptual features 
with the {\sl much more familiar} step-up in complexity from the triangles' (2, 3) to the quadrilaterals' (2, 4). 

\m 

\n{\bf Remark 5} This completes realization of the {\it qualitatively-distinct triplets} of $N$-Body Problems for values of $N$ 
(1, 2, 3) in 1-$d$, 
(2, 3, 4) in 2-$d$, 
(3, 4, 5) in 3-$d$, 
(4, 5, 6) in 4-$d$ ... and 
\be
(d, d + 1, d + 2) \m 
\ee 
in general dimension $d$.  

\m 

\n On the other hand, $(d, N) = (4, 6)$ is revealed to be substantially more of a sequel to $(d, N) = (3, 5)$ than $(d, N) = (3, 6)$ is.
Such sequels are moreover never-ending, by Proposition 3's formula. 
These constitute the first parallel above the basisland diagonal, i.e.\ the minimal (linear) dependentlands \c{Minimal-N}.

\section{0-$d$ case}\label{0-d}

\n Let us first consider $d = 0$.
In this case, 
\be 
\mL \, \in \, SO(0) = id  \m \Rightarrow \m  \mL = 1            \m . 
\ee 
Also while $\bK \, \in \, SO(n)$, $d = 0$ supports no vectors for these internal rotations to act on in the first place. 

\m 

\n This leads to the following simplifications.  
Firstly,  
\be 
\mJ \,  \bK^{\sT}  =  \mL \, \mJ 
                   =         \mJ                                    \m . 
\ee 
Secondly, 
\be 
\mJ = \mL \, \Lambda \, \bH^{\sT}                                         \m .  
\ee
collapses via $\mL \, \in \, SO(0)$ being a zero-dimensional vector and $\Lambda$ a $0 \times n$ matrix to just  
\be 
\mJ = 0                                                 \m . 
\ee 
Consequently,  
\be 
Isot(\mR) = SO(n)                                      \m . 
\ee 
Our working moreover {\sl nominally} returns this as the sole possibility, for all that we argued that the $SO(n)$ action is trivial in the first place. 

\m 

\n The corresponding orbit geometry is given by, introducing our notation  $\FrO(d, r)$, 
\be 
\FrO(0, \, 0)  \:=  \frac{SO(n)}{SO(n)} 
                   \es  id 
		 		    =   \{pt\}                               \m .    
\ee 
$n = 1$, i.e.\ $N = 2$ minimally realizes this.

\section{1-$d$ case}\label{1-d}

Next let 
\be
\mL \, \in \, SO(1) \m \Rightarrow \m  \mL = 1                    \m , 
\ee 
\be 
\bK \, \in \, SO(n)                                               \m , 
\ee  
so again 
\be 
\bJ \, \bK^{\sT}  =  \mL \, \bJ 
                  =  \bJ                                      \m . 
\label{1}
\ee 
$d = 1$ is moreover minimal as regards supporting class distinctions.    
The collinear case is invariant under 
\be 
\bK = id
\ee 
alone, while the maximal collision is invariant under arbitrary $\bK$.  

\m 

\n $\bJ$'s principal decomposition 
\be
\bJ = \mL \, \bLambda \, \bH 
\label{2}
\ee 
now simplifies by (62) and $\bLambda$ taking the form of a $1 \times n$ matrix, i.e.\ a row vector 
\be 
(\lambda_1 \, | \, 0 \, . \, . \, . \, 0 \, )   \m .
\ee 
\n  The following cases are then supported, as indexed by rank.  

\m 

\n {\u{Case 0)}} $r = 0$ corresponds to the maximal collision.

\m

\n {\u{Case 1)}} $r = 1$ corresponds to the linear shape. 

\m 

\n In either case, 
\be 
\Lambda  =  (\Sigma \, | \, 0)                   \m .  
\ee 
\be 
\bA \, \in \, O(1) \m \Rightarrow \m \bA  =  \pm 1
\ee  
or absent 
\be 
\bB  \, \in \, O(n - 1)                        \m . 
\ee    
So 
\be
\bC  \mma  \bD \m \mbox{ are trivially 1, absent in some order}  \m .  
\ee 
\n We next require (42-44) to hold

\m 

\n {\u{Case 1)}} Set $r = 1$. 
Then there is no $\bD$. 
So 
\be 
\mbox{det}(A) = \pm 1  \m \Rightarrow \m  A = \pm 1  \m ,
\ee 
forming $C_2$. 
Thus 
\be
\mbox{det}(\bB)  =  1                                \m , 
\ee 
by which 
\be 
\bB  \, \in \, O(n - 1)                                     \m .
\ee  
So
\be 
Isot(\mR) = O(n - 1)                                   \m . 
\ee 
\n{\u{Case 0)}} Set $r = 0$.  
\be 
\mbox{dim}(\bA) = 0  \m \Rightarrow \m  \bK = \bB          \m . 
\ee 
As 
\be 
\mbox{det}(\bB) = 1                                    \m ,
\ee 
the $\bB$'s form $SO(n)$. 
Thus 
\be 
Isot(\mR) = SO(n)                                      \m . 
\ee 
We summarize these results in Fig \ref{7-Isot}.a) including also the continuous parts of each group in Fig \ref{7-Isot}.b) 
 
\m 

\n The corresponding orbit geometries are given by, using the notation  $\FrO(d, r)$, 
\be 
\FrO(1, 0)  \:=  \frac{SO(n)}{SO(n)} 
                \es  id 
		 		 =   \{pt\}                               \m ,  
\ee 
\be 
\FrO(1, 1)  \:=  \frac{SO(n)}{O(n - 1)} 
                \es  \mathbb{RP}^2                        \m .   
\ee 

\n Proposition 1 follows from the following series of coincidences. 

\m 

\n $SO(p)$ undefined for $p < 0$ and $SO(0) = \emptyset$ removes all isotropy groups for $N = 1$ and all but the top one for $N = 2$.

\m 

\n $N = 3$ already has the $C$-generic number of distinct isotropy groups in 1-$d$, as $SO(2)$ and $C_2$.

\section{2-$d$ case}\label{2-d}

Finally let 
\be  
\bL \, \in \, SO(2)                                          \m .
\ee
and 
\be  
\bK \, \in \, SO(n)                                          \m .
\ee 
We are to solve 
\be 
\bJ  \, \bK^{\sT} = \bL \, \bJ                               \m , 
\label{R-K-Q}
\ee 
with $\bJ$ moreover admitting the principal decomposition 
\be 
\bJ = \bL \, \Lambda \, \bH                                  \m , 
\label{L-Lambda-H}
\ee 
where 
\be 
\bL \, \in \,  SO(2)                                         \mma  
\bH \, \in \,  SO(n) 
\ee 
and $\Lambda$ is a $2 \times n$ matrix of form
\be 
\mbox{\Huge(} 
\s{\mbox{$\lambda_1$}}{0} \m \s{\mbox{0}}{\lambda_2} \, \, \mbox{\Huge |} \s{\mbox{0 \m . \m . \m .}}{\mbox{0 \m . \m . \m .}} \s{\mbox{0}}{0}  \mbox{\Huge)}  \m . 
\label{Lambda}
\ee 
Without loss of generality, 
\be 
\lambda_1 \m \geq \m       0                                              \mma 
\lambda_1 \m \geq \m  |\lambda_2|                                         \m .
\ee
By Sec \ref{main}'s argument, it suffices to take 
$\bK \, \in \, SO(n)$
and 
$\mL \, \in \, SO(2)$ 
satisfying 
\be 
\mL \, \Lambda \, \bK^{\sT} = \Lambda                                         \m . 
\ee 
2-$d$ supports three cases,  indexed by rank as follows. 

\m 

\n{\u{Case 0)}} $r = 0$: the maximal collision.

\m 

\n{\u{Case 1)}} $r = 1$: linear shapes. 

\m 

\n{\u{Case 2)}} $r = 2$: generic planar shapes

\m 

\n In this case, the uniformative version of Sec \ref{main}'s working holds.  

\m

\n{\u{Case 2)}} There is no $\mD$ block, so we require  
\be 
\mbox{det}(\bA) =  1                                                        \m . 
\ee 
On the other hand, for $r < 2$, there is an orthogonal matrix such that 
\be 
\mbox{det}(\bD) = \mbox{det}(\bA)                                             \m , 
\ee  
so once again the uninformative version of Sec \ref{main}'s working holds.

\m 

\n For $r = 2$, 
\be 
\mbox{det}(\bA) = 1                                                         \m ,
\ee 
so Sec \ref{main} gives that 
\be 
\mbox{det}(\bB) = 1
\ee 
as well. 
Thus 
\be 
\bB \, \in \, SO(n - 2)                                                          \m ,
\ee 
independently of $\bA$.  

\m  

\n $\bA$ can moreover take values such that 

\m 

\n {\u{Subcase i)}}  `Asymmetric planar top' configurations,  or 
\be
\lambda_1 \neq \lambda_2                                                 \m , 
\ee 
\n{\u{Subcase ii)} `symmetric planar top' shapes   
\be 
\lambda_1 = \lambda_2                                                    \m .
\ee 
\n We interpret i) and ii) in shape-theoretic terms in the Conclusion.  

\m 

\n In Subcase i), setting
\be 
\bA  \es  \mbox{\Huge(}  \s{\mbox{\normalsize$\ia_1$ \m $\ia_2$  }}{  a_3  \m \, \, a_4  }    \mbox{\Huge)}                                                                 \m , 
\ee 
\be 
a_2(\lambda_2 - \lambda_1) = 0                                                                                                             \m , 
\ee 
since  
\be 
\lambda_1 \m \neq \m \lambda_2 \mma a_2 = 0                                                                                                \m .
\ee 
Also 
\be 
a_2(\lambda_2 - \lambda_1) = 0                                                                                                             \m , 
\ee 
so 
\be 
a_3 = 0                                                                                                                                    \m .
\ee   
Thus 
\be
\bA  \es  \mbox{\Huge(} \s{\mbox{$a_1$}}{0} \m \s{\mbox{0}}{a_4} \mbox{\Huge)}                                                              \m .  
\ee 
But $\bA$ is orthogonal, so 
\be
\bA  \es  \pm \mbox{\Huge(} \s{ \mbox{1}}{0} \m \s{\mbox{0}}{\m 1}   \mbox{\Huge)} \m \mbox{ or }                                                              \m ,
\ee 
i.e.\ the $\bA$'s form $C_2$. 
Thus we have 
\be 
Isot(\mR)      \es       C_2 \times SO(n - 2) 
           \m \cong \m   O(n - 2)                                                                                              \m . 
\ee 
\n In subcase ii), 
\be 
\bA \lambda_1 \bI  =  \lambda_1 \bI \bA                                                                                                     \m , 
\ee 
so there is no restriction on $A$ [just like for 3-$d$'s case iii) \cite{ML00}]. 
So the $\bA$'s in this case form $SO(2)$, and  
\be 
Isot(\mR)      \es      SO(2)     \times SO(n - 2) 
           \m \cong \m  O(n - 2)                                                                                                            \m . 
\ee
\underline{$r = 1$ case} 
\be 
\bA \, \in \, O(1)
\ee 
so, as the determinant condition gives no restriction, 
\be 
\bA = \pm 1                                                                                                                                 \m .
\ee   
\n Then 
\be 
\pm 1 = \bA 
      = \mbox{det}(\bB)                                                                                                                     \m ,
\ee 
so 
\be 
\mB       \, \in \,   O(n - 1) 
\ee 
unrestrictedly, and so 
\be 
Isot(\bR)      =      O(n - 1)                                                                                                              \m . 
\ee
\n{\u{$r = 0$ case}}  
\be
\mbox{dim}(\bA) = 0  \m \Rightarrow \m  \mK = \bB                                                                                           \m .
\ee 
As 
\be 
\mbox{det}(\bB) = 1                                                                                                                         \m , 
\ee 
the $\mB$'s form $SO(n)$, so 
\be 
Isot(\bR) = SO(n)                                                                                                                         \m . 
\ee 
We summarize these results in Fig \ref{7-Isot}.a), including also the continuous parts of each group in Fig \ref{7-Isot}.b).  

\m

\n The corresponding orbit geometries are given by  
\be 
\FrO(2, \, 0)  \es  \frac{SO(n)}{SO(n)} 
                   \es  id 
				    =   \{pt\}                                            \m :
\ee
real-projective spaces 
\be 
\FrO(2, \, 1)  \es  \frac{SO(n)}{O(n - 1)} 
                   \es  \mathbb{RP}^{n - 1}                               \m . 
\ee 
Grassmann spaces
\be 
\FrO(2, \, 2)  \es  \frac{SO(n)}{SO(2) \times SO(n - 2)} 
                \es  \w{G}(n, \, 2)                                       \m ,  
\ee
and $C_2$ quotients of Stiefel spaces 
\be 
{\cal O}(2, \, 2, \, \lambda_1 \neq \lambda_2)  \es  \frac{SO(n)}{O(n - 2)} 
                                                \es  \frac{S(n, \, 2)}{C_2}                               \m . 
\ee 
The continuous counterparts are points 
\be 
\FrO_+(2, \, 0)  =  \{pt\}                 \m ,
\ee
spheres   
\be 
\FrO_+(2, \, 1) = \mathbb{S}^{n - 1}       \m ,
\ee
Grassmann spaces 
\be 
\FrO_+(2, \, 2) = \w{G}(n, \, 2)         \m , 
\ee  
and Stiefel spaces 
\be 
\FrO_+(2, \, 2, \, \lambda_1 \neq \lambda_2) = S(n, \, 2)         \m . 
\ee 
\n Proposition 1's 2-$d$ case follows from the following series of points. 

\m 

\n The end of the preceding section's working applies again; 
$SO(p)$ undefined for $p \leq 0$ and $SO(0) = \emptyset$ removes all isotropy groups for $N = 1$ and all but the top one for $N = 2$.

\m 

\n In 2-$d$, the first and third isotropy subgroups are conflated by $SO(0) \times SO(2)$ collapsing to  $SO(2)$, whereas the fourth is knocked out.  

\m 

\n $N = 4$ has the $C$-generic number of full isotropy groups in 2-$d$, as $SO(3)$, $O(2)$, $SO(2) \times SO(2)$ and $C_2$.

\section{Lattice of isotropy subgroups}

I furthermore observe a sense in which Mitchell and Littlejohn's condition for $N = 5$ is not generic.  
This is based on considering the {\sl bounded lattice formed by} the isotropy subgroups; 
the continuous parts for this are presented for $d = 2$ and 3 in Figs \ref{Isot-Latt}.a) and \ref{Isot-Latt}.b).\footnote{This might in general be just a bounded poset, 
but in all cases featuring in the current article, it is a fortiori a bounded lattice.} 
%
{\begin{figure}[ht]
\centering
\includegraphics[width=0.85\textwidth]{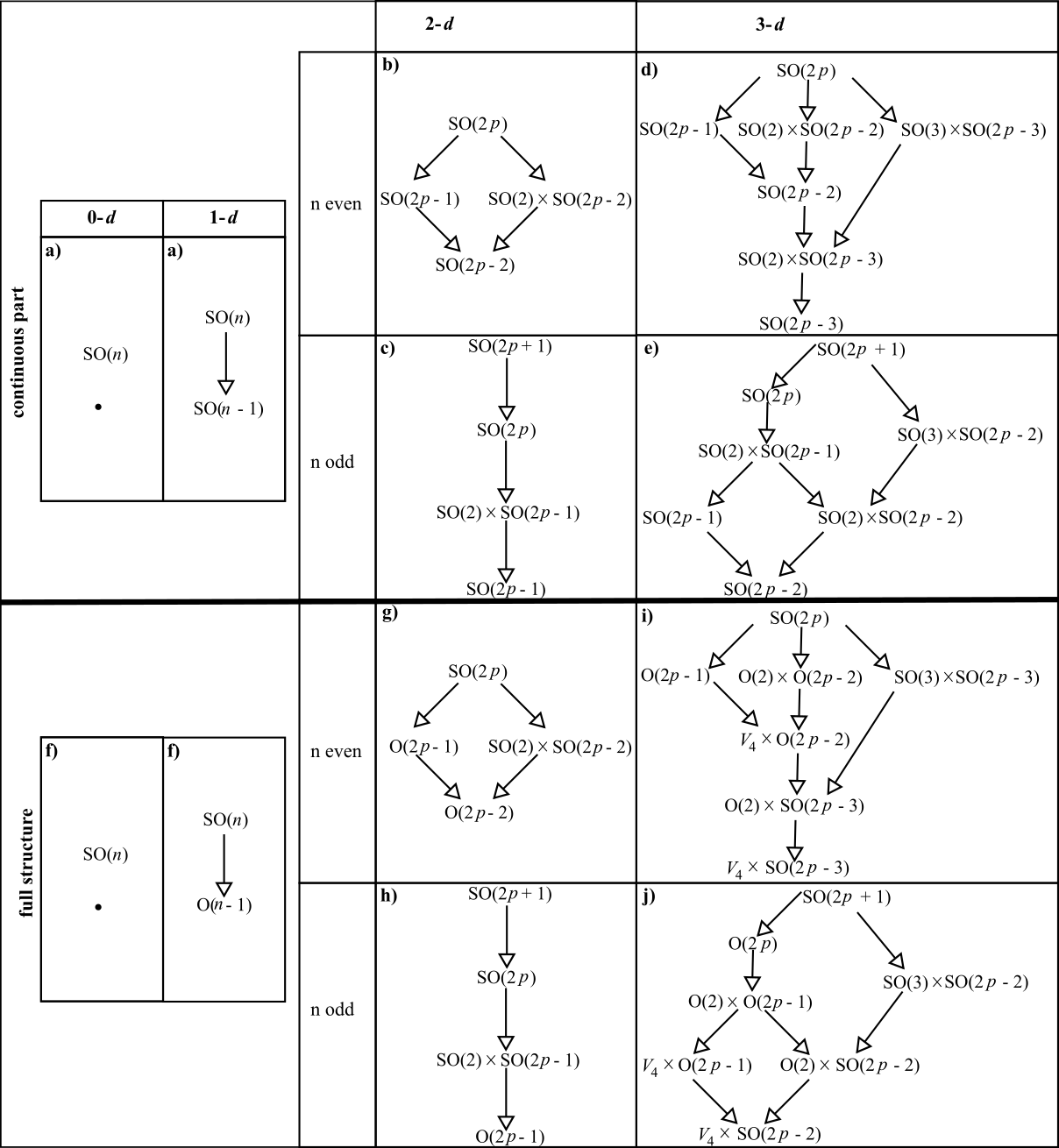}
\caption[Text der im Bilderverzeichnis auftaucht]{\footnotesize{The general lattices.
}} 
\label{Target-Lattices}\end{figure} } 
%
{\begin{figure}[ht]
\centering
\includegraphics[width=0.8\textwidth]{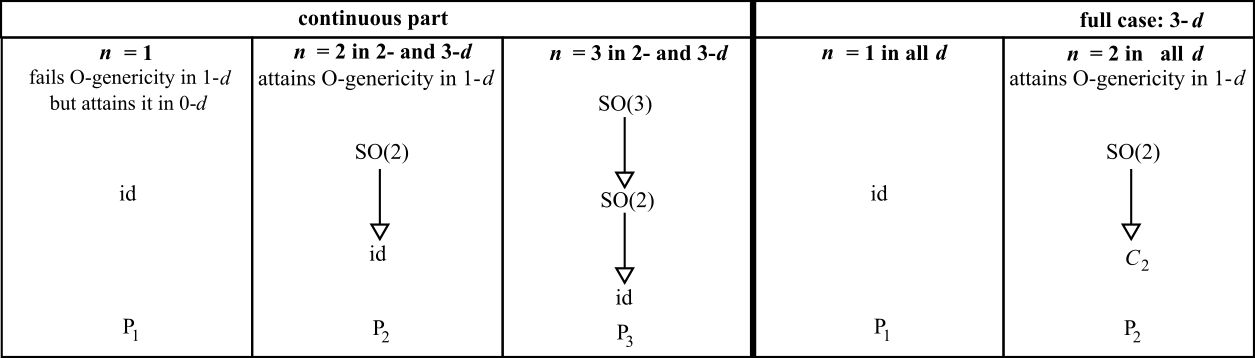}
\caption[Text der im Bilderverzeichnis auftaucht]{\footnotesize{The exceptional cases' $k$-chains $P_k$ of isotropy subgroups in 0-, 1- and 2-$d$}} 
\label{Isot-Latt}\end{figure} } 

\m 

\n The even--odd distinction of these lattices in 2- and 3-$d$ follows from the number of Casimirs going up by one for every even $SO(n)$ 
                                                                                                                   but not at all for every odd $SO(n)$.  
Thus 
\be 
SO(2\,p - 1) \m \leq \m  SO(2 \, p) 
\ee 
leaves one Casimir unused, which can be used to generate an extra $SO(2)$, so 
\be 
SO(2) \times SO(2\,p - 1) \m \leq \m  SO(2 \, p)                                  \m .  
\ee 
On the other hand, 
\be 
SO(2\,p - 2) \m \leq \m SO(2 \, p - 1) 
\ee
uses up all of the Casimirs, so an extra $SO(2)$ subgroup cannot be included.  

\m

\n We can place a sequence of qualitative criteria in terms of increasing complexity of the bounded lattice of isotropy subgroups as follows.   

\m 

\n For arbitarary $d$, $N = 2$'s isotropy subgroup lattice is only a point.   

\m 

\n $N = 3$'s isotropy subgroup lattice is the first with a distinct top and bottom but has no middle.
In 1-$d$, this attains genericity.  

\m 

\n $N = 4$'s isotropy subgroup lattice is the first to have a middle but is still just a chain. 

\m
   
\n $N = 5$'s isotropy subgroup lattice is the first with a nontrivial -- rather than just chain -- middle.    

\m 

\n{\bf Proposition 3} Realizing the generic lattice of the continuous parts of the isotropy subgroups requires 
$N = 4$ in 2-$d$ and 
$N = 8$ in 3-$d$.    

\m 

\n{\bf Proposition 4} 
\n i)   For $d \neq 3$, 
\be 
n = 2 \, d \m \mbox{ ( i.e. \ $N = 2 \, d + 1$ ) }  
\ee
is an upper bound (`B-genericity', with `B' standing for bounding) on O-genericity.   

\m 

\n ii)  For $d = 3$, $n = 7$ is required. 

\m 

\n We need $d = 0$, $n = 0$, $d = 1$, $n = 2$ and $d = 2$, $n = 4$.
The 3-$d$ case's exceptionality is rooted in the 
\be 
so(4) \m = \m so(3) \times so(3)
\label{Acci}
\ee 
accidental relation.

\section{Conclusion}

We consider orbits and isotropy groups for $N$-body problems in 0, 1 and 2-$d$. 
This complements \cite{ML00, Minimal-N}'s coverage of the 3-$d$ case, 
particularly in the light of \cite{Minimal-N, Minimal-N-2}'s detailed study of $N$'s interplay with $d$.  
For Mitchell and Littlejohn's \cite{ML00} notion that we term C-genericity -- `C' for `counting' -- 2, 3 and 4 Body Problems are minimal in dimensions 0, 1 and 2.  
For the Author's notion of O-genericity (`O' for order-theoretic, referring to the bounded lattice of isotropy subgroups), 2, 3 and 5 Body problems are minimal in these dimensions. 
This 2 is an exception to the $N = 2 \, d + 1$ rule, due to having only one object to order providing a more stringent bound.  
3 is moreover also an exception by the Lie group accidental relation \ref{Acci}. 
We also tabulated the isotropy groups and orbit topologies and geometries for our small-$N$ Body Problems in dimensions 0, complementing \cite{ML00} 
and making use of \cite{Minimal-N}'s observation that Stiefel and Grassmann spaces occur in bottom and top roles.  

\m 

\n Applications include the following. 

\m 

\n 1) that some of the increase in complexity in passing from 3 to 4 and 5 body problems in 3-$d$ is already present in the more-well known setting 
of passing from intervals to triangles and then to quadrilaterals in 2-$d$. 

\m 

\n 2) That not (3, 6) but (4, 6) is a natural theoretical successor of (3, 5); however, we leave this, and other higher-$d$ $N$-Body Problems for subsequent occasions.  

\m 

\n 3) Such consideration isotropy groups and orbits is moreover a model for a larger case of interest, namely that of GR's reduced configuration spaces.  

\m 

\n We finally comment that, at the level of shapes, for 2-$d$ triangles the equal-eigenvalue case is regular to the non-equal eigenvalue case being irregular = tall-or-flat.  
{\it Regular} here means that the mass-weighted partial moments of inertia of the base and median are equal \cite{FileR, III}.
Without mass weighting, these are in the proportion found in the equilateral triangle. 
The 2-$d$ quadrilaterals are moreover characterized by consideration of pairwise regularities in the subsystems formed by ignoring one of the three Jacobi separations 
that the quadrialteral's frame supports \cite{QuadI, I, Minimal-N}.

\m 

\n{\bf Acknowledgments} I thank Chris Isham and Don Page for concrete discussions about configuration space topology, geometry and background independence. 
I thank Don, Jeremy Butterfield, Malcolm MacCallum, Enrique Alvarez and Reza Tavakol for support with my career.


\end{document}